# Notions of Equivalence in Software Design

**Résumé**

Les méthodes de conception dans des systèmes d'information créent souvent des descriptions de logiciel en utilisant des langages formels. Néanmoins, la plupart des concepteurs de logiciel préfèrent décrire le logiciel en utilisant des langages naturels. Cette distinction n'est pas simplement une question de pragmatisme. Les langages naturels ne sont pas identiques à des langages formels; en particulier, langages naturels ne suivent pas les notions de l'équivalence employées par langages formels. En cet article, nous montrons l'existence et la coexistence de différentes notions de l'équivalence en un extension de la notion des oracles utilisés dans langages formels. Ceci permet à des distinctions d'être faites entre les oracles dignes de confiance assumés par des langages formels et les oracles non dignes de la confiance employés par des langages naturels. En examinant la notion de l'équivalence, nous espérons encourager les concepteurs de logiciel à repenser l'endroit de l'ambiguïté dans la conception de logiciel.

**Mots clefs:**
Program, logiciel, équivalence, langage formel, langage naturel.

**Abstract**

Design methods in information systems frequently create software descriptions using formal languages. Nonetheless, most software designers prefer to describe software using natural languages. This distinction is not simply a matter of convenience. Natural languages are not the same as formal languages; in particular, natural languages do not follow the notions of equivalence used by formal languages. In this paper, we show both the existence and coexistence of different notions of equivalence by extending the notion of oracles used in formal languages. This allows distinctions to be made between the trustworthy oracles assumed by formal languages and the untrustworthy oracles used by natural languages. By examining the notion of equivalence, we hope to encourage designers of software to rethink the place of ambiguity in software design.

**Key-words:**
Programs, Software, Equivalence, Formal language, Natural Language.

*David KING*
Research Student
University of York
Department of Computer Science,
+44 1904 432749
**dk@cs.york.ac.uk**

*Chris KIMBLE*
Lecturer
University of York
Department of Computer Science,
+44 1904 433380
**kimble@cs.york.ac.uk**

# Preface

Language is clearly important as a medium for communication in software design. Nonetheless, discussions of language in software design have been dominated by debates about the role of formal languages. For over 40 years, formal languages have been held as the ideal medium for communication software design. Formal languages are precise with clear definitions and elegant characterisations. Formal languages then, should be ideal for software design (Zemanek H, 1985). Yet, despite over 40 years of argument, an overwhelming majority of software design remains heavily reliant on natural languages. Why?

In this paper, we propose that natural languages and formal languages have different, and fundamentally incompatible, notions of equivalence. We accept the notion of equivalence used in formal languages is universally applicable to all program designs, however this does not mean the same notion of equivalence is applicable to all software designs. Consequently, while ambiguity must be eliminated for descriptions of programs, we will argue that ambiguity has a rightful place in software design.

# 1. The importance of equivalence

We begin this paper with a working definition of equivalence: equivalence is simply taken to mean an isomorphic relationship between the things on either side of an equivalence relationship. That is to say, that things can be mapped between the one side of the relationship and the other without changing their meaning. This notion of equivalence will be examined in detail in § 2.

## 1.1 Programs or software?

Although the term's 'software' and 'program' are often treated as though they were equivalent, most researchers now draw a distinction between 'software design' and 'program design'. In the early years, the problems faced by designers were focused on the machine, for example, describing how a particular calculation should be carried out. All that was needed to solve these problems was to work out the exact sequence of instructions, the program, which could be fed into a machine. Thus, program design is firmly centred on the computer.

However, as problems became larger designers quickly realised that the machine-based focus did not always help when it came to solving 'real world' problems. For example, a business might regularly need to know the stock levels held in a warehouse. From the point-of-view of the program designers, it does not matter much whether the results need to be returned by stock number, level of stock, or location of the stock. However, it does matter to the business. Having to manually re-sort a program output by stock number to work out the location of the stock can be done, but it significantly reduces the usefulness of the program. Gradually these descriptions of the 'larger-purposes' of the program became known as software. Thus while software design includes program design, software designers must look beyond the computer for their craft.

In this paper, we will concentrate on the problem of software design. However, before moving onto problems of software design, we need to clarify the meaning of certain other terms.

## 1.2 Definitions

As a starting point for our definitions, we begin with a recognition that an unperceived universe exists: a universe beyond the reach of any individual designer. This unperceived universe we will call *reality*, defined as "*...that which underlies and is the truth of appearances or phenomena*" (Simpson J. and Weiner E. 1989). In common with other philosophical discussions, the emphasis in discussions on reality focused on what really exists, rather than the existence perceived by any individual designer (Audi R. 1995). The perceptions of an individual designer can be important to the success of the software design (Sommerville I. and Sawyer P. 1997) as only by capturing these viewpoints can a designer fully understand the needs of all the users (Graham T. 1996).

Thus in software design, in addition to '*unperceived reality*' we need a notion of '*perceived reality*'; we will call this perceived reality the *representation*. The representation can be thought of as the mental model of an individual designer; characterising their perceived reality. The representation can therefore be defined as "*[t]he operation of the mind in forming a clear image or concept*" (Simpson J. and Weiner E. 1989).

However, these mental models are not directly transmissible. Instead, some intermediate form is needed; a form we will call the *description*. We will discuss descriptions in detail in §2. For now we will simply define descriptions as "*[a] statement which describes, sets forth, or portrays*" (Simpson J. and Weiner E. 1989).

Having presented some working definitions for the terms descriptions, representation and reality, we will now begin to define the terms relating these concepts. We will start with the relationship between reality and a representation. Essentially, this relationship characterises the representation as the result of the reception of stimuli from reality. Following Charles Meadow and Weijing Yuan (1997), we will refer to this as *information*. By contrast, the formation of a description from a representation is a conscious act by the designer who knowingly chooses the form of the description. We will therefore refer to this relationship as *knowledge*. All five terms are briefly summarised pictorially in Figure 1.

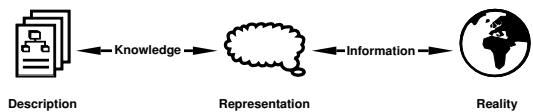

**Figure 1 - Summary of Definitions and Terms**

## 1.3 Equivalence in software design

Although researchers in software design have often used programming theory as an inspiration for their work, programming theory is bound to a strict notion of equivalence. This formal notion of equivalence conflates several more general notions that may otherwise held to be distinct, thus making adequate definitions of some terms impossible.

Within formal theory, the perceptions of an individual designer are irrelevant. Formal theory focuses instead on uncovering an objective reality: one untainted by the views of the individual. Consequently, formal theory cannot naturally distinguish between reality and representations. Formal theory does not permit valid representations in an individual mind to diverge from reality; any mental models designer has must map seamlessly onto a larger reality.

Likewise, for formal theory any valid description must have a seamless, isomorphic, mapping to the representation. Consequently, there must be some way of exactly characterising any individual designer's representation. Without this constraint, forming descriptions with unambiguous meanings is impossible.

This means a formal theory can deal with written descriptions of reality, as the assumption is that the written description, the mental representation and the underlying reality are all equivalent. Thus, formal theory makes no practical distinction between the concepts of description, representation and reality.

In program design, the advantages of formal theories heavily outweigh the disadvantages of not being able to make a distinction between description, representation or reality. However, for software design, the reverse is true; software designers often need to make such distinctions.

## 1.4 Equivalence and Epistemology

Traditionally, the ascendance of formal theory has meant software designers borrowed extensively from rationalist epistemological arguments. Broadly, for the rationalist epistemological position, "reason is the source of all knowledge". Hence, "*[i]f reason is the source of all knowledge, then everything that can be known including the natural world must be intelligible and rationally explicable*" (Jack A. 1993). Formal equivalence is important to rationalist arguments because of the emphasis on objective truth. Once an objective truth has been found, then other equivalent truths must also be true. Consequently, a few truths and a network of equivalence arguments can be used to construct detailed descriptions.

Thus, using a formal notion of equivalence strongly favours the use of rationalist arguments. However, moving from the formal to other notions of equivalence allows software designers to consider other epistemological positions. For example, as we will see in §4, object-oriented software design methods need a notion of "equivalent until further notice". Such a notion is difficult to construct using rationalist arguments but is natural in another style of epistemological argument: empiricism. The empiricist school argues that knowledge is the result of observation and experience. Thus, an empiricist would expect a notion of truth to be along the lines of "experienced to be true" or "observed to be true". This need not be an eternal truth; empiricist arguments are responsive to truths that change in the light of experience.

The difficulty of establishing eternal truths using empiricist arguments means that formal notions of equivalence cannot be used. This lack of a formal notion of equivalence has led researchers to find ways of treating design methods more naturally inclined to empiricism (such as object-oriented design) in a purely rationalist manner.

By extending the notion of equivalence, we hope to encourage software designers to consider other epistemological arguments that may form a more appropriate foundation for their work than rationalist ones. We will begin by examining other notions of equivalence such as those that exist in natural languages before moving onto a more detailed definition of equivalence in §2.2.

## 2. Equivalence in software

### 2.1 Equivalence in natural languages

For many researchers in Information Systems, the key attraction of formal languages is that they are unambiguous. All formal languages are capable of creating crisp, elegant descriptions with properties chosen by the language designer. However, the crispness and elegance of formal languages has prejudiced many software design methods against ambiguity. Ambiguity in descriptions of software design is seen as detrimental to the ability of the description to communicate the intent of the designer. In arguing for a more general notion of equivalence, we are arguing for the retention of ambiguity in descriptions of software.

We will begin our examination of equivalence by exploring two components of language: written descriptions and spoken descriptions. Although in a formal language written and spoken descriptions would be synonymous, this is not the case for natural languages. For natural languages (Bloomfield L. 1935, pp. 21):

"*Writing is not language, but merely a way of recording language by means of visible marks ...*"

Natural languages can be used to communicate without first needing to create a fixed and immutable set of rules. Instead, spoken and written descriptions evolve independently. Yet, despite the independent evolution of written and spoken forms of language, some relationship between the two is preserved. We contend this ability is given by a more general notion of equivalence than the one accepted by formal languages.

Natural languages need to retain their ambiguity because their component descriptions are not underpinned by a single set of rules. Instead, the components of the language change at different rates as the language changes. In contrast, the components of formal languages cannot change independently because the rules governing the language cannot change. Change in formal languages therefore creates new languages whereas in natural languages the history of change becomes part of the language itself.

## 2.2  What is equivalence?

Having highlighted these differences between natural and formal languages, we now return to the question "what is equivalence?" According to the Oxford English Dictionary, equivalence means (Simpson J. and Weiner E. 1989):

"*Equal in value. Now only in more restricted uses: (a) of things regarded as mutually compensating each other, or as exchangeable*"

The key notion here is of exchangeability. For descriptions, this means that equivalent descriptions can be exchanged without affecting the sense of either description, that is we could say that two descriptions are equivalent if they both appeal to the same mental model, as shown in Figure 2.

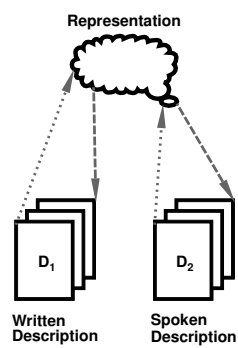

**Figure 2 - A naïve view of equivalence**

By appealing to the same mental model, two descriptions ought to appeal to the same meaning. Either description captures the intended meaning, so either description could be used to characterise it. For example, a designer ought to have the same mental model for both spoken and written descriptions of the same concept. In this case, the spoken and written descriptions are equivalent. Either the spoken or the written description could be used without affecting the meaning of the description.

However, we noted previously that representations are not directly transmissible. This creates a problem: we cannot guarantee the mental model used by the designer is the same for both written descriptions and spoken descriptions. Without a direct means of examining the representation, we cannot establish the substance (or even the number) of mental models used by the designer.

In essence, then, descriptions appeal to representations for meaning. Let us now examine this appeal in more detail. The first step is that a description claims to characterise some feature of the representation. If this claim is upheld, we can move onto the next stage: comparing the representations. If the representations compare, we can then deduce that, since the two claims are valid, the two descriptions must be equivalent. The problem for the naïve view of equivalence in Figure 2 is that without being able to directly compare representations, we cannot complete the second stage of the appeal.

Nonetheless, we cannot even reach the second stage without first being able to verify the claim of a description to characterise the representation. Let us therefore put an oracle between a description and the representation. The task of this oracle is simply to adjudicate on the claim "This description means …".

Ignoring the details of the oracle, and assuming the representation as the input, our problems at the first stage of the appeal are then twofold. Firstly, how do know that the oracle always refers to the same mental model when answering a question? Without a guarantee that a description always points to the same mental model, we have no guarantee that the meaning is preserved and, without this, we have no guarantee that two descriptions are equivalent. Secondly, even assuming the oracle is always using the same mental model, how do we know the answers the oracles provide are repeatable? Equivalent descriptions should always be exchangeable, if not we might run into the situation where two descriptions are exchangeable one moment but not exchangeable at some later point.

However, although these are difficult problems, both will only occur if we insist the oracle is trustworthy, that is (1) the oracle must always give the same output given the same input, and (2) the oracle must divulge what that input was in order to be able to demonstrate its trustworthiness. We will return to the consequences of these conditions in §3. For the moment, we will consider untrustworthy oracles further.

Untrustworthy oracles are almost the reverse of trustworthy ones. Firstly, untrustworthy oracles may give different outputs for the same inputs. Secondly, untrustworthy oracles do not state what their inputs are - only the outputs. Untrustworthy oracles can answer the question "This description means …", however, the questioner will not know the grounds on which the question is an-swered nor will they know if they would get the same answer again for the same input.

We can now use this notion of an untrustworthy oracle to describe a more general notion of equivalence. To do this we will insert an untrustworthy oracle between the description and the representation, as shown in Figure 3. As before, the oracle will adjudicate on the question "This description means …".

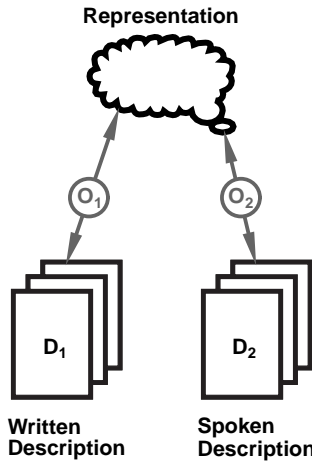

**Figure 3 - A rigorous foundation for equivalence**

Each description has its own untrustworthy oracle. In Figure 3 the written description, $D_1$ has the untrustworthy oracle $O_1$ and the spoken description $D_2$ has the untrustworthy oracle $O_2$. Given that both $O_1$ and $O_2$ are untrustworthy, we can ignore the representation (i.e. the input), leaving Figure 4.

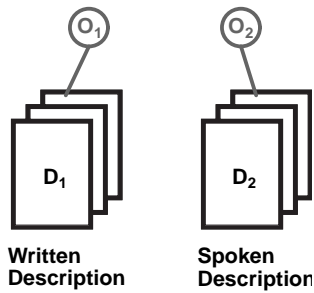

**Figure 4 - Focusing on the description**

This then leaves two descriptions with two untrustworthy oracles, giving a general scheme that can be used as the foundation of a theory of equivalence. Rather than detailing this scheme here, the next two sections will concentrate on four specific cases that are directly applicable to current approaches to software design. The details of the application of this theory to software design can be found in the accompanying paper, King D. and Kimble C. (2004)

## 3. Equivalence in Rationalist Arguments

Rationalist arguments concentrate on demonstrations of truth. In rationalist arguments it is not enough to claim truth: truth must be shown by some means. The require-ment for demonstration in rationalist approaches favours an emphasis on the description, or more usually on systems of descriptions. A system is simply a founding theory, from which other descriptions can be created. The ideal for rationalist approaches is self-describing systems. Self-describing systems are closed, so the system not only describes a truth in descriptions, but also describes what truth is and how it can be recognised. As the rules of the system governing exchanges are fixed, or immutable, we can therefore claim that if two descriptions are equivalent by the conventions of the system, they are immutability equivalent. In this paper, we denote *immutable equivalence* using the symbol '≡.' We can now represent an immutable equivalence between written and spoken descriptions as shown in Figure 5.

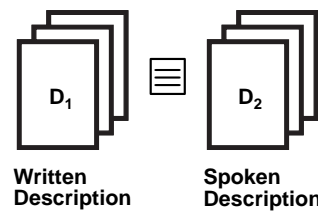

**Figure 5 - Immutably equivalent descriptions**

However, closed systems are rare and usually also quite trivial (Claude C. and Paun G. 2000). More often, such systems either end up containing descriptions that are both simultaneously true and false; or descriptions whose truth cannot be proved. Demonstrating the presence of an immutable equivalence between two descriptions therefore requires an appeal to something beyond the system. Usually these appeals are to self-evident truths, or axioms, that can be accepted as true without debate. Given a set of axioms, other truths can then be constructed within the system.

For example, we could create an axiom stating, "This description is provable …" using two related systems of descriptions, such as logic and a sub-set of number theory. Within each system, we can create an exact definition of what we mean by "This description is provable …". In the system of logic, we could use Alan Turing's mathematical machines and in number theory, we could use Alonzo Church's recursive functions. Neither of these systems is perfectly closed, but both have an exact definition of what we mean by the statement "This description is provable …". Further, because the two systems are related, we can create descriptions in one system having an exact analogue in the other. For example, Turing machines (logic) have an exact analogue using Church's recursive functions (number theory). We now have two systems, each containing the axiom "This description is provable…" along with a precise meaning. Further, we can translate descriptions between the two systems because the two systems are related. Therefore, instead of arguing over whether the axiom "This description is provable…" in logic is exactly the same as the axiom "This description is provable…" in number theory, we simply define the two axioms to be equivalent. This allows the creation of immutable descriptions in the related systems, as long as the oracles attesting to the truth of the founding axioms are trustworthy.

Using the symbol '□' for trustworthy oracles, these conclusions are shown in Figure 6. By following this series of assumptions, systems that are almost completely closed can be created. These systems are more usually known as **formal** systems.

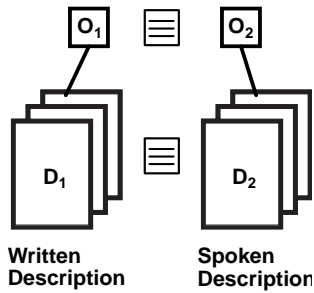

**Figure 6 - A Formal Notion of Equivalence**

However, the formal style of reasoning only works when the oracles can be assumed trustworthy as, without trustworthy oracles, immutable equivalences within the system might fail to hold outside the system.

The possibility of untrustworthy oracles poses many problems in software design. In many software designs, simply assuming properties of the description are also properties of reality (or the representation) is sometimes a step to far. Michael Jackson gives the following example taken from the design of an aircraft control system (Jackson M. 1996).

One requirement of this system was:

```
REVERSE_ENABLED if and only if MOVING_ON_RUNWAY
```

The designers concluded that the following axioms held:

```
WHEEL_PULSES_ON if and only if WHEELS_TURNING

WHEELS_TURNING if and only if MOVING_ON_RUNWAY
```

Thus, the designers concluded that the following immutable equivalence held:

```
WHEEL_PULSES_ON if and only if MOVING_ON_RUNWAY
```

In making this conclusion, the designers assumed that an immutable equivalence in the description also held for things outside the description. If the designers had access to a trustworthy oracle, they could be sure that this equivalence actually did hold: unfortunately, they did not. Instead, they had to rely on their own experience (an untrustworthy oracle) to judge whether the axioms always characterised the domain in question.

In this case, the designers experience proved inadequate, they did not realise that if the wheels aquaplaned on a wet runway, `MOVING_ON_RUNWAY` held but `WHEELS_TURNING` did not. Although the developers could form an immutable description inside the description, without a trustworthy oracle, they could not guarantee that it held outside.

The situations faced by the designers in this example are common and are often tackled by semi-formal methods of software design. **Semi-formal** design methods allow for the formation of immutable equivalence between descriptions, but the designers have to be aware that these equivalences are based on untrustworthy oracles. Consequently semi-formal methods of design must employ some means of checking whether the immutable equivalence between descriptions hold outside the description.

Using the symbol 'O' to represent untrustworthy oracles, we can now redraw Figure 6 as shown in Figure 7. This forms a new notion of equivalence, held by all semi-formal methods of software design. Note that in Figure 7 no immutable equivalence between the two oracles can be assumed, because both oracles are untrustworthy.

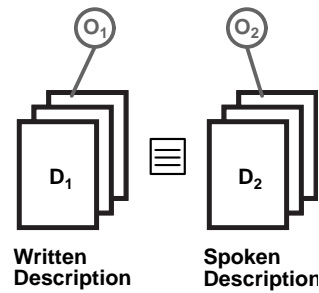

**Figure 7 - Semi-Formal Notion of Equivalence**

Having now examined two approaches to software development based on rationalist epistemological arguments, we will turn to two approaches to software development based on empiricist arguments.

## 4. Equivalence in Empiricist Arguments

Basing the conclusion of the design methods on untrustworthy oracles presents semi-formal design methods with something of a problem. No semi-formal method can guarantee that properties claimed from the description hold in the real world. Consequently, although semi-formal design methods such as structured design were popular in the early days of software design, more recent design methods have attempted to infer properties of the description from reality, rather than trying to infer properties of reality from the description. In essence, these are design methods based on an empiricist epistemological position.

From an empirical viewpoint, an appeal to experience can be used as an external check on any claim; the immutability of our set rules can always be checked against reality. We could therefore argue that experience forms the foundations of a trustworthy oracle. As long as descriptions are carefully constructed from the demonstrated experience of designers, there should be no doubt over the claims of any individual description to characterise reality.

This approach leaves us with a problem. By working from reality, we cannot form immutable equivalencies between descriptions, as the final proof of equivalence lies not in the properties of the description, but in the properties of reality. Hence, the truth of any claimed equivalence can only be based on previous experience. While two descriptions may be equivalent from previous experience, there is no guarantee that some future experience will not invalidate rules forming the description.

This is exactly the problem faced by **object-oriented** software design methods, which all choose to argue from the experience of reality. Object-oriented design methods assume trustworthy oracles, using reality as the final arbitrator of claims made by descriptions. An immutable set of equivalences therefore exists between all things in reality. However, even if the oracles are trustworthy, the existence of immutable rules for reality does not prevent the future experience of designers from having to change the description, as illustrated by the previous example.

As we noted in §1.4 equivalencies in object-oriented designs are in the form of "equivalent until further notice". We will call this form of equivalence *mutable equivalence*, because no systematic conventions can be formed using only the properties of the description. The rules governing exchangeability for mutable equivalences are not guaranteed: a mutable equivalence may hold at one point in time but not another.

We represent the concept of mutable equivalence using the symbol '≡'. Since object-oriented design methods assume the presence of trustworthy oracles, the notion of equivalence assumed by these methods can be shown in Figure 8.

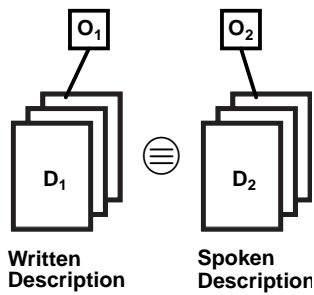

**Figure 8 - The Object-Oriented Notion of Equivalence**

In the preface, we said that we would argue that ambiguity has a rightful place in software design: using mutable equivalence with trustworthy oracles is about as far as most software design methods have gone. Nonetheless, some design methods have questioned the assumption of trustworthy oracles. Object-oriented design methods assume that, because reality is independent of the experience of any individual designer, a trustworthy oracle (an appeal to an objective observer) must exist. However, does everybody share the same experience of reality?

A few software design methods, such as Peter Checkland's soft systems methodology, focus the problems of reaching a consensus on, for example, who or what a 'user' might be or what a 'payroll system' is and what it might do. These methods argue that far from assuming all designers think alike, a design method must take into account the individual prejudices, experience and inclinations of the participants (Checkland P. 1981).

However, removing reality as an external arbitrator also removes the assumption that trustworthy oracles underpin a description; now we must assume that all equivalencies are mutable and all assumptions are open to challenge. This fluid notion of equivalence is used in what we have called **holistic** design methods. Unlike object-oriented methods, equivalencies are open to challenge not only from future experience, but from past and current experience as well. Substituting untrustworthy oracles for the trustworthy oracles in Figure 8 produces Figure 9 showing the notion of equivalence used in holistic design methods.

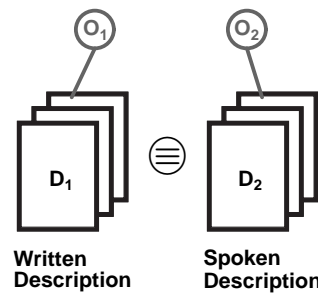

**Figure 9 - The Holistic Notion of Equivalence**

# 5. Conclusion

Traditionally, most software design methods have drawn inspiration from program design. However, methods of program design are underpinned by a systematic theory capable of characterising every valid description forming a program. This theory is so well formed that, for all practical purposes, the assumptions made by its founding axioms are irrelevant for most program designs. However, once we move beyond the program in a machine, the questions around its founding axioms move sharply back into focus. Software design does not have a systematic theory for characterising every possible software description. This calls into question two key assumptions used by formal theories in defending equivalencies.

The first is assumption is that axioms (descriptions) are formed though the actions of a trustworthy oracle. With a trustworthy oracle, the relationship between a description and either the representation or reality is maintained. Experience in software design has shown that this assumption cannot always be defended. Often establishing the relationship between the description and the representation (or the representation and reality) is a major problem for software designers. Even if a relationship has been found, it is often open to challenge; software design methods must be capable of dealing with its untrustworthy oracles.

The second assumption made by formal theories is that by using an established body of theory we can defend immutable equivalences between descriptions. However, software has no all-encompassing body of theory. Instead, equivalencies must often be defended from experience rather than theory. Equivalence defended from experience we have called mutable equivalence, and some software design methods must be prepared to use this form of equivalence.

Evidence from recent empirical studies (King D. and Kimble C. 2004) suggests that no one form of equivalence is universally suited to all software designs. Further, this evidence suggests that even within a single software design process, the notion of equivalence used by software designers changes.

Based on these observations and the arguments presented in this paper, we reject any attempt to create a unified theory of software design and instead advocate a move towards using 'ad hoc' theories of software design. Here we use the term ad hoc to signify "*devoted, appointed to or for some particular purpose*" (Simpson J. and Weiner E. 1989). We believe that each notion of equivalence represents the core assumption of a particular school of software design. Arguments in software design must now move away from discussions of which school is right and instead each school should concentrate on producing evidence outlining where a particular notion of equivalence is, and is not, most appropriate.

Such discussions are unlikely to produce clear, unambiguous conclusions. Nonetheless, human designers are naturally good at coping with ambiguity. Let us therefore turn this ability into a strength instead of condemning it as a weakness.